\documentclass[times]{article}

\usepackage{moreverb}
\usepackage{subcaption}		
\usepackage{graphicx}

\newcommand\BibTeX{{\rmfamily B\kern-.05em \textsc{i\kern-.025em b}\kern-.08em
T\kern-.1667em\lower.7ex\hbox{E}\kern-.125emX}}

\begin{document}

\title{Cold Object Identification in the Java Virtual Machine}

\author{Kim T. Briggs\footnote{IBM Canada, 770 Palladium Drive, Ottawa, ON, Canada, E-mail: briggs@ca.ibm.com}, 
	Baoguo Zhou\footnote{Faculty of Computer Science, University of New Brunswick, Fredericton, E3B 5A3, N.B., Canada, E-mail: barry.zhou@unb.ca}, 
	Gerhard W. Dueck\footnote{Faculty of Computer Science, University of New Brunswick, Fredericton, E3B 5A3, N.B., Canada, E-mail: gdueck@unb.ca}}




\maketitle

\begin{abstract}
Many Java applications instantiate objects within the Java heap that are persistent but seldom if ever referenced by the application. Examples include strings, such as error messages, and collections of value objects that are preloaded for fast access but they may include objects that are seldom referenced. This paper describes a stack-based framework for detecting these ``cold" objects at runtime, with a view to marshaling and sequestering them in designated regions of the heap where they may be preferentially paged out to a backing store, thereby freeing physical memory pages for occupation by more active objects. Furthermore, we evaluate the correctness and efficiency of stack-based approach with an Access Barrier. The experimental results from a series of SPECjvm2008 benchmarks are presented.
\end{abstract}

For submission to `Software: Practice and Experience'

\section{Introduction}
Long-running Java applications~\cite{Gosling:2013:JLS:2462622}, such as web servers and network security monitors, may preload and retain large numbers of objects within the Java heap~\cite{lind99} in order to allow fast application access at runtime. In many cases, some of these objects are infrequently referenced by the application. We refer to these objects, which are persistent in the Java heap but seldom referenced, as cold objects. 

The presence of cold objects in the heap is problematic insofar as they may be collocated in virtual memory~\cite{JavaMemory} with more active objects. Any page of virtual memory that contains all or part of a cold object may also contain parts of more active objects, and application references to the active objects will prevent the page from being swapped out of virtual memory. As a result, large applications that commit most or all of the available physical memory may experience undue memory pressure. 
Additionally, active objects and cold objects are co-located, when active objects are accessed, both active objects and cold objects might be loaded into cache at the same time, actually cold objects will not be accessed. Therefore, cold objects will degrade the cache-hit performance. 

If cold objects are collected and moved to cold regions, both page fault and cache-miss performance could be decreased. Furthermore, cold regions can be excluded from Garbage Collection (GC)~\cite{lind99},
if they contain only leaf objects and primitive arrays. Therefore, pause times caused by the GC can be reduced as well.

As the cold area becomes populated with cold objects, operating system primitives such as madvise()~\cite{Love:2007:LSP:1205435} may be used to inform the operating system that pages mapped to the cold area may preferentially be swapped out of resident memory, thereby freeing those pages for occupation by more active objects. The cold area can then be monitored, for example, with continuous reference sampling and periodic calls to mincore(), to detect when presumed cold objects become active and to take appropriate action. 

Management of cold objects is most relevant in the context of long-running applications with large heap requirements. For that reason, the balanced garbage collection~\cite{IBMGCPolicies, IBMBalancedGCPolicy, IBMBalancedGC} (GC) framework was selected as a basis for investigating cold object management. The balanced collector improves application performance by partitioning the heap into a large number of regions of equal size and limiting the number of regions that are included for collection in partial GC cycles. This reduces the frequency of more time-consuming global GC cycles, which involve all heap regions.

In this paper, we present a stack-based framework to identify cold objects. Cold objects  have been identified and harvested successfully in many SPECjvm2008~\cite{SPECjvm2008,SPECjvm2008Performance} applications. At the same time, we evaluate the correctness and efficiency of stack-based solution with an Access Barrier mechanism~\cite{BarrierConcept}. All experiments are performed on IBM's J9 Virtual Machine (JVM)~\cite{IBMJ9}.

\section{Stack-Based Cold Object Identification Framework}
A stack-based framework is used to identify and harvest cold objects. The framework supports identification and collection of cold objects. The main idea of the measurement of cold objects is to periodically walk thread-local stacks and mark active references. 
Whenever a Java method is invoked, a new stack frame will be generated. Since local variables and passed arguments~\cite{Gosling:2013:JLS:2462622} will be stored in the stack frame, object references corresponding to local variables and passed arguments in the current stack frame are considered to be active. 
After a period of time when no new active objects are being discovered, subtraction of the collection of objects found to be active from the collection of live objects reveals the collection of live, inactive (cold) objects. Once they have been identified, cold objects can be harvested form the main heap and sequestered in a designated cold area.

\subsection{Cold Region Reservation and Instrumentation}
When the JVM~\cite{lind99} starts, a preset number of contiguous regions are reserved for the cold area. Cold regions are excluded from copy forward and compaction during partial GC~\cite{IBMBalancedGC} cycles, except to receive objects that have been identified as cold and copied from pinned regions, as described below.  

In order to preclude the need to traverse cold regions during mark/sweep~\cite{MarkingAlgorithm, generationalGCInventor} actions, only arrays of primitive data (eg, char[]) and leaf objects (objects with no reference-valued fields) are considered as collectible cold objects. This constraint, in conjunction with the leaf object optimization feature available in the IBM Java virtual machine~\cite{IBMJ9}, ensures that objects within cold regions can be correctly marked but not touched by the marking scheme. This constraint can be relaxed to include Java objects that contain only primitive-valued fields.

Objects that have been sequestered in cold regions are monitored for reference activity. To that end, each cold region is instrumented with an activity map containing one bit per potential object location within the region, as for the collector's mark map. The stacks of all mutator threads are periodically walked to collect active heap references. Any mutator reference to an object within a cold region is marked by setting the corresponding bit in the activity map. 

The number and total size of objects that are sequestered in the cold area and the incidence of activity involving these objects are the main outcomes of interest for this paper.

\subsection{Pinned Region Selection and Instrumentation}
When marking active objects on the heap, since objects might be moved by copy-forward or compaction~\cite{abua04} of the GC, some regions are selected to be pinned. Pinned regions are excluded from partial GC collection sets~\cite{lind99} sets. That means objects in the pinned region will not move, which facilitates tracking and marking active objects.

The balanced garbage collector~\cite{IBMBalancedGCPolicy} assigns a logical age to each heap region, which reflects the number of GC cycles that the contained objects have survived. Allocations of new objects occur in the youngest regions (age 0), and persistent objects are progressively copied forward into increasingly older regions until they are copied into tenured regions (age 24). In the balanced GC, tenured regions are excluded from partial GC collection sets.

In order to enable detection of cold objects, a number of tenured regions are first pinned so that they are excluded from partial GC collection sets~\cite{IBMBalancedGC} cycles. This ensures that objects contained within pinned regions maintain a fixed location within the mutator address space. Pinned regions are also instrumented with activity maps to record which objects have been sampled from mutator stacks. Cold objects can be identified within a pinned region, after a preset amount of time T$_{cold}$ has elapsed since the most recent setting (0 $\rightarrow$ 1) of an activity bit, by subtracting the activity map from the mark map.

\subsubsection{Pinned Region Selection.}
The region pinning framework partitions the regions of the balanced GC heap into four collections:
\begin{enumerate}
\item Young regions (age $<$ 24)
\item Unpinned regions (age 24, not pinned)
\item Pinned regions (age 24, pinned)
\item Cold regions (the cold area)
\end{enumerate}

Unpinned regions are considered for pinning at the end of each partial GC cycle. They are selectable if they have an allocation density $d$ (ratio of allocated bytes to region size $R$) exceeding a preset threshold $D_{hi}$. Additionally, the total size of potentially collectible cold objects contained in the region must be greater than $0.01R$. Selectable regions are ranked at the end of each partial GC cycle according to the region pinning metric value $P$ that reflects the volume of activity in each region:

\begin{equation}
   P = mma(r)*d   
\end{equation}
where $r$ is the number of mutator references to contained objects since the end of the previous partial GC cycle, $r$ reflects the object activity in the region since any reference that is found on the stack frame is considered to be active. The $mma(r)$ is the modified moving average of $r$ with a smoothing factor of 0.875:

\begin{equation}
   mma(r_{0} )=0;  mma(r_{n} ) = \frac{(7*mma(r_{n-1} )+ r_{n})} {8},  n>0  
\end{equation}

The maximum number of regions that may be pinned at any time is determined by a preset parameter P$_{max}$. At the end of every partial GC cycle, if the number of currently pinned regions $n$ is less than $P_{max}$, up to $P_{max}$ – $n$ selectable regions may be pinned. 

Two strategies for selecting regions for pinning were implemented. The pinning strategy is determined by a JVM parameter that is interpreted when the JVM starts and remains fixed while the mutator runs. With selective pinning, only the most active selectable regions are pinned. An active selectable region must satisfy $mma(r) > r > 0$ and $sum(r) > R$, where $sum(r)$ is the sum of $r$ over all previous partial GC cycles. The average pinning metric value $P_{avg}$ from the collection of all selectable regions with non-zero pinning metric value is computed and only regions satisfying $P > P_{avg}$ are selectable for pinning. The activity maps for these regions should converge more quickly and cold object identification should be more accurate after a period ($T_{cold}$) of quiescence. 

The alternative pinning strategy, unselective pinning, pins in decreasing order of pinned metric value selectable tenured regions up to a preset maximum ($P_{max}$). Cold objects will be found only in tenured regions that persist in the heap. Unselective pinning should converge to a pinned region collection that contains all of these regions.  

In either case, pinned regions are unpinned, at the end of every partial GC cycle, if their density falls below a preset low density threshold $D_{lo}$, the total mass of eligible objects (primitive arrays) falls below $0.01R$, or they survive a period of inactivity $> T_{cold}$ and the contained collectible cold objects are moved into the cold area. 

All pinned regions are unpinned at the start of a global GC cycle, or when the cold area becomes full. Pinning is resumed after the global GC completes or when space becomes available in the cold area.

\subsubsection{Pinned Region Instrumentation.}
When a region is pinned, it is instrumented with an activity map to track reference activity within the region. The activity map contains an activity bit for each mark bit. Activity bits are initially 0 and are set when a reference to the corresponding object is sampled. The region is also walked to assess the number of marked objects $n_{marked}$, the number of marked collectible objects $n_{collectible}$, and the respective total sizes $m_{marked}$ and $m_{collectible}$ of these collections. 

Three timestamps are maintained to record the time $t_{pinned}$ at which the region was pinned, the time $t_{inactive}$ of the most recent setting of an activity bit, and the time $t_{walked}$ that the region was most recently walked. Pinned regions are walked, and $t_{walked}$ is updated to the current time t, whenever $t - t_{walked} > T_{cold}/ 4$. Current values for $n_{marked}$, $m_{collectible}$, $m_{marked}$, $m_{collectible}$, and $d$ are obtained each time the region is walked.

Over time, the rate at which activity bits are set will diminish, until a period of time $> T_{cold}$ has elapsed with no new activity bits set. The collectible cold objects in this region can then be identified and copied into the cold area.

\subsubsection{Mutator Thread Instrumentation.}
The primary sources for reference collection are the mutator stacks, which are periodically walked down from the top-most frame until a frame that has not been active since the most recent stack walk. Frame equality is determined on the basis of the frame base pointer and a hash of the stack contents between the frame base and stack pointers. Each mutator thread is instrumented with a fixed-length buffer for reference collection and two arrays of stack frame traces--one to hold traces from the most recent stack walk and one to hold traces from the current stack walk. 

References from each active frame are added to the mutator's reference buffer. Stack walks are discontinued if the reference buffer overflows (collected references are retained for activity map updates). If the stack frame buffer for the current stack overflows, the mutator continues to walk the stack and collect references while matching current frame base pointer and hash against the previous stack until a match is found. The next previous stack is then composed from the head of the current stack and the tail of the previous stack. Any missing frames between these stack segments are of little consequence--if the next stack walk continues past the end of the head segment it will fall through to the tail segment and eventually find a match, and setting the activity bits for redundant samples collected from frames with missing traces is  idempotent. 

In addition, two timestamps $w_{start}$ and $w_{end}$ are maintained for each mutator thread to record the start and end times of the most recent stack walk. References collected from stack walks started before the most recent GC cycle are discarded. 

\subsubsection{Activity Sampling Daemon.}
When the JVM is initialized, a thread activity sampling daemon is started to control reference activity sampling when mutator threads are executing and to harvest references collected by mutator threads.
 
The daemon thread remains in a paused state during GC cycles. Between GC cycles the daemon interacts with mutator threads by polling each active mutator thread at approximately 1 millisecond intervals. During each polling cycle, the daemon instruments each previously uninstrumented mutator thread and signals it to start a stack walk. It harvests collected reference samples from previously instrumented mutator threads that have completed a stack walk and signals these threads to start a new stack walk. Mutator threads receive these signals and commence the stack walk at their safe point~\cite{jone05}.

For each harvested reference sample the daemon increments the reference activity counter $r$ for the containing region (young, pinned, unpinned, or cold). Additionally, if the referenced object is contained in a pinned or cold region, the daemon sets the corresponding bit in the region'��s activity map. No explicit synchronization is required to set or test the activity bits, since they are set only on the daemon thread and tested only on the master GC thread, and these threads never access region activity maps concurrently.

\subsubsection{Cold Object Collection.}
The region pinning framework attempts to pin a collection of tenured regions that contains as much of the mutator's active working set as possible. This may seem counterintuitive, given that we are attempting to identify persistent objects that are almost never in the working set. Cold object identification is like looking for shadows in a windowless room---they are easier to see when the lights are turned on. Pinning the most active regions is expected to reduce the likelihood of identifying as cold objects that are actually just dormant in the context of current mutator activity. For pinned regions that are receiving few active references all or most objects would be identified as cold under a fixed $T_{cold}$ threshold.

In the presence of high reference activity, the activity map of a pinned region can be expected to converge more quickly to a stable state where no new activity bits are being set.  After a fixed time T$_{cold}$ has elapsed since the last change to the state of the activity map, the pinned region will be included in the copy forward collection set for the next partial GC cycle, if the pinned region has an accurate remembered set card list and no critical regions in use. When a pinned region is included in the copy forward collection set, collectible cold objects are copied into the next available cold region while all other objects are copied into other unpinned regions. After all objects have been copied out the region is unpinned. 

Cold regions are instrumented as for pinned regions in order to allow reference activity to be tracked. At present, all objects that are copied to cold regions remain in the cold area, without compaction or copying, until they are collected as garbage or the mutator terminates.

\section{JVM PROFILING CONFIGURATIONS}
All JVMs were compiled with gcc version 4.4.7 with optimization enabled (-O3). Two JVMs were produced for profiling:

\begin{enumerate}
\item linux: generic JVM with no reference sampling.
\item ssd-stack: stack sampling enabled.
\end{enumerate}

The JVM run configurations are shown in (Table \ref{tab_parms}).
\begin{table}[!h]
\caption{Running parameters}
\begin{center}
\begin{tabular}{| p{1.8cm} | p{2.6cm} | p{1.5cm} | p{2.0cm} | p{1.5cm} |}
\hline
JVM	&Pinning strategy&JIT&Run time(s)&Tcold(s)
\\
\hline\hline
linux&none&	enabled&9600 x 2&900\\
\hline
ssd-stack&selective&enabled&9600 x 2&900\\
\hline
ssd-stack&unselective&enabled&9600 x 2&900\\
\hline
\end{tabular}
\end{center}
\label{tab_parms}
\end{table}

The linux and ssd-stack JVMs were each run twice with the Just-in-Time~\cite{IBMJIT,JITConcept} compiler (JIT) enabled. The only data of interest from the second runs were the SPECjvm2008 benchmark scores, which were the basis for comparison of overall performance of the linux versus ssd-stack JVMs. The ssd stack JVM was run in two modes -- selective or unselective pinning -- in order to permit comparison of cold object identification between these region pinning strategies. 

Four SPECjvm2008~\cite{SPECjvm2008} benchmarks (compiler.compiler, derby, xml.transform, xml.validation) were selected for profiling. The linux JVM was executed twice for each benchmark. The ssd-stack JVMs ran each benchmark twice with selective pinning and twice with unselective pinning. Each profiling run executed a single iteration of one benchmark for 9600 seconds. Sampling interval is 1 $ms$, and T$_{cold}$ is 15 minutes. 

All profiling runs were performed on a 1.8GHz 8 Core/16 Thread (Xeon) server running CentOS version 6.4. No other user applications were active during any of the profiling runs.

\section{RESULTS}
The first three JVM run configurations from Table \ref{tab_parms} (linux, ssd-stack/selective pinning, ssd-stack/unselective pinning) were used to determine the runtime heap characteristics of each SPECjvm2008 benchmark and to allow performance comparisons between the linux and ssd-stack JVMs. 
Heap characteristics most salient to cold object identification with activity tracking are the numbers of tenured regions and the distributions of mutator activity within young, unpinned, and pinned regions.

\subsection{SPECjvm2008 Scores versus Linux}
\begin{figure}[ht!]
  \centering
  
  \begin{subfigure}{0.45\textwidth}
    \includegraphics[width=\textwidth]{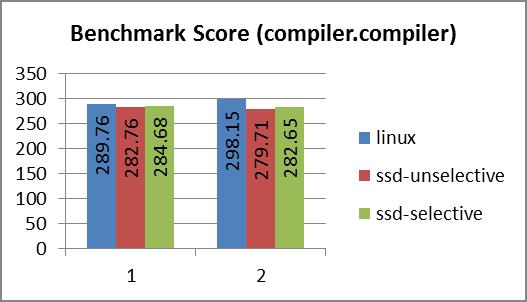}
    \caption{Compiler.compiler score}
    \label{fig:Compiler.compilerscore}
  \end{subfigure}
  \begin{subfigure}{0.45\textwidth}
    \includegraphics[width=\textwidth]{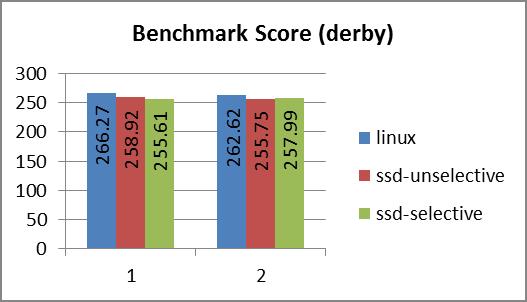}
    \caption{Derby score}
    \label{fig:Derbyscore}
  \end{subfigure}

    \begin{subfigure}{0.45\textwidth}
    \includegraphics[width=\textwidth]{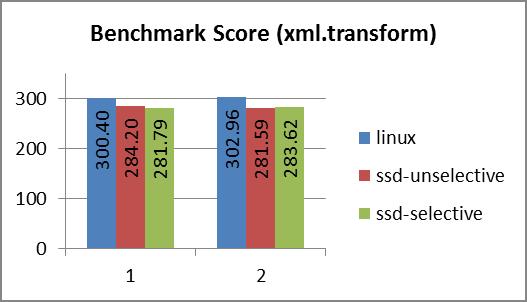}
    \caption{xml.transform score}
    \label{fig:xml.transformscore}
  \end{subfigure}
  \begin{subfigure}{0.45\textwidth}
    \includegraphics[width=\textwidth]{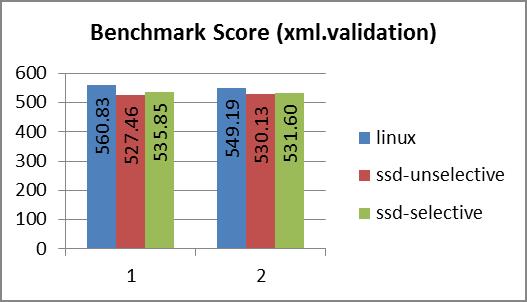}
    \caption{xml.validate score}
    \label{fig:xml.validatescore}
  \end{subfigure}
  
  \caption{Running performance of the ssd-stack JVMs}
  \label{fig_perform}
\end{figure}

Runtime performance of the ssd-stack JVMs (selective and unselective pinning) versus the linux JVM was assessed using SPECjvm2008~\cite{SPECjvm2008} scores for two runs of each benchmark: compiler.compiler, derby, xml.transform, and xml.validation. The resulting benchmark scores are plotted in Figure~\ref{fig_perform}.

Performance degradation was calculated as the ratio of the difference between the linux and ssd-stack scores to the linux score. The overall average performance degradation for the ssd stack JVMs was 0.04 (4\%). Performance degradation ratios versus linux for all runs of the ssd stack JVMs are listed in Table \ref{tab_overhead}.

\begin{table}[!h]
\tiny
\caption{Overhead caused by the feature of cold objects}
\begin{center}
\begin{tabular}{| p{3.2cm} | p{1.2cm} | p{1.2cm} | p{1.2cm}| p{1.2cm}|}
\hline
Pinning strategy&\multicolumn{2}{|c|}{selective}&\multicolumn{2}{|c|}{unselective}\\
\hline
Benchmark run&1&2&1&2\\
\hline
compiler.compiler&0.02&0.05&	0.02&0.06\\
\hline
derby&0.04&	0.02&0.03&0.03\\
\hline
xml.transform&0.06&0.06&0.05&0.07\\
\hline
xml.validation&0.04&0.03&0.06&0.03\\
\hline
\end{tabular}
\end{center}
\label{tab_overhead}
\end{table}

\subsection{Garbage Collection Metrics versus Linux}
Summary garbage collector metrics are shown in Table~\ref{tab_GC_metrics} for all benchmark runs with the linux JVM and the ssd-stack JVM with unselective and selective pinning. Compaction times were significantly lower for all of the ssd-stack JVM runs with the compiler.compiler benchmark. Considering the average total GC times for each pair of runs, compiler.compiler had slightly better total GC times for ssd-stack unselective and selective pinning compared to linux (4\% and 2\% lower, respectively), as did xml.transform (4\% and 5\%). Average total GC times were slightly worse for derby (4\% and $<$1\% higher than linux) and xml.validation (2\%, 2\%).  Much of the ssd stack GC overhead is incurred in compiling pinned region and cold collection statistics and streaming them to a printable log file. These statistics are informative only and can be suppressed to reduce overhead.


\begin{table}[!h]
\caption{GC metrics (in $ms$)}
\begin{center}
\footnotesize{
\begin{tabular}{| l | r | r | r | r | r | r |}
\hline\hline
compile.compile&	Compact&	 Copyforward&	Glb. mrk&	Incr. mark&	Sweep&	Total\\
\hline\hline
linux-r1&244& 	 3,517,715& 	 33& 	 1,166,619& 	 54,731& 	 4,739,401\\
\hline
linux-r2&	 339& 	 3,442,073& 	 38& 	 1,157,656& 	 55,105& 	 4,655,271\\
\hline
unselective-r1&	 31& 	 3,401,260& 	 35& 	 1,112,495& 	 52,078& 	 4,565,965\\
\hline
unselective-r2&	 99& 	 3,314,364& 	 31& 	 1,088,048& 	 51,770& 	 4,454,374\\
\hline 
selective-r1&	 35& 	 3,445,347& 	 38& 	 1,128,614& 	 52,869& 	 4,626,962\\
\hline 
selective-r2&	 37& 	 3,405,924& 	 33& 	 1,112,630& 	 51,835& 	 4,570,519\\
\hline 
derby&	Compact&	Copyforward&	Glb. mrk&	Incr. mark&	Sweep&	Total\\
\hline\hline
linux-r1&12,147& 	 587,707& 	 40& 	 2,884& 	 2,552& 	 605,641\\
\hline
linux-r2&8,252& 	 581,722& 	 35& 	 2,916& 	 2,051& 	 595,282\\
\hline 
unselective-r1&12,998& 	 608,385& 	 36& 	 2,856& 	 2,569& 	 627,179\\
\hline
unselective-r2&13,222& 	 598,226& 	 34& 	 2,768& 	 2,507& 	 617,069\\
\hline 
selective-r1&10,879& 	 570,283& 	 38& 	 3,030& 	 2,483& 	 587,038\\
\hline 
selective-r2&10,000& 	 598,301& 	 44& 	 2,860& 	 2,217& 	 613,744\\
\hline 
xml.transform&	Compact&	Copyforward&	Glb. mrk&	Incr. &	Sweep&	Total\\
\hline\hline
linux-r1&	 37& 	 375,820& 	 40& 	 761& 	 342& 	 377,126\\
\hline 
linux-r2&	 31& 	 376,827& 	 39& 	 773& 	 327& 	 378,127\\
\hline 
unselective-r1&	 30& 	 362,166& 	 35& 	 799& 	 328& 	 363,486\\
\hline
unselective-r2&	 31& 	 357,149& 	 32& 	 814& 	 360& 	 358,513\\
\hline
selective-r1&	 31& 	 359,097& 	 32& 	 847& 	 362& 	 360,498\\
\hline
ssd-selective-r2&	 39& 	 353,823& 	 34& 	 813& 	 372& 	 355,211\\
\hline
xml.validation&	Compact&	Copyforward&	Glb. mrk&	Incr. mark&	Sweep&	Total\\
\hline\hline
linux-r1&	 35& 	 707,832& 	 33& 	 872& 	 283& 	 709,135\\
\hline
linux-r2&	 97& 	 726,825& 	 38& 	 952& 	 306& 	 728,299\\
\hline
unselective-r1&	 33& 	 742,721& 	 41& 	 611& 	 236& 	 743,722\\
\hline
unselective-r2&	 36& 	 725,050& 	 29& 	 786& 	 269& 	 726,251\\
\hline
selective-r1&	 30& 	 730,070& 	 39& 	 958& 	 292& 	 731,468\\
\hline
selective-r2&	 35& 	 725,577& 	 36& 	 907& 	 274& 	 726,909\\
\hline
\end{tabular}}
\label{tab_GC_metrics}
\end{center}
\end{table}

\subsection{Region Age and Activity}
\begin{figure}[ht!]
  \centering
  
  \begin{subfigure}[b]{0.45\textwidth}
    \includegraphics[width=\textwidth]{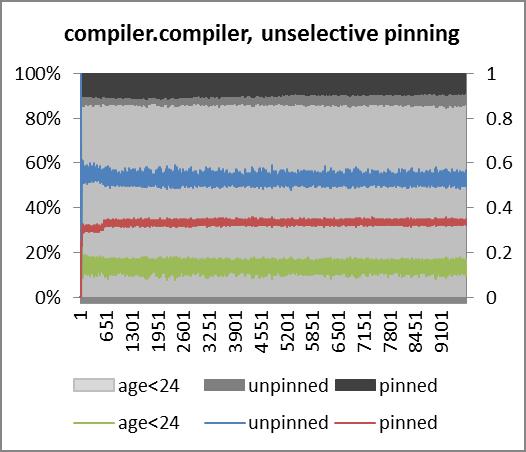}
    \caption{Compiler.compiler, Unselective Pinning}
    \label{fig:2a}
  \end{subfigure}
  \begin{subfigure}[b]{0.45\textwidth}
    \includegraphics[width=\textwidth]{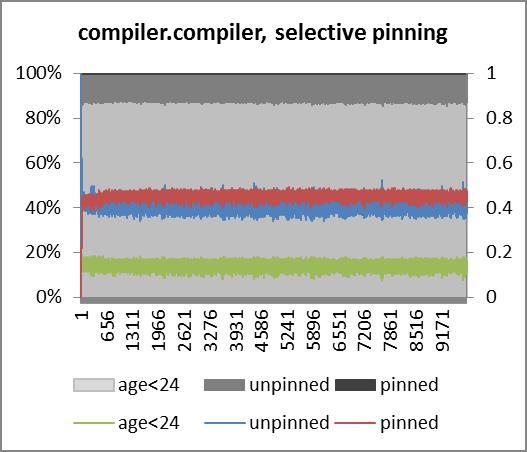}
    \caption{Compiler.compiler, Selective Pinning}
    \label{fig:2b}
  \end{subfigure}
  
  \caption{Compiler.compiler Activity}
\end{figure}

Figures \ref{fig:2a}-\ref{fig:2b} represent the relative region counts (greyscale, left axis, percentage of total marked region count) and reference counts (colored lines, right axis, proportion of total reference count) for the young, unpinned, and pinned parts of the heap after each partial GC cycle. Plots are presented for unselective and selective pinning for each benchmark executed with the ssd stack JVM, using data collected from the first of two runs. Partial GC cycle counts are represented on the horizontal axes.

The compiler.compiler benchmark (Figures \ref{fig:2a}, \ref{fig:2b}) was atypical in that it showed a predominance of young regions (over 90\% of marked regions) that receive a relatively high proportion (almost 0.2) of reference activity. The other benchmarks (Figures 3a $-$ 5b) all showed a predominance of unpinned and pinned regions that receive almost all reference activity. Regardless of pinning strategy, they also showed a tendency for an initially high concentration of reference activity within pinned regions that diminished over time. This is not surprising since both pinning strategies favor selection of regions receiving high reference activity. The activity maps of pinned regions with high reference activity tend to converge relatively quickly. Most regions are pinned early in the mutator lifecycle and remain pinned until they are cold collected, their contents become dereferenced, or the mutator ends. When they are cold collected the remaining objects, active or not collectible, are redistributed to other unpinned regions, so that active objects tend to become more diffusely scattered over time. 

\begin{figure}[ht!]
  \centering
  
  \begin{subfigure}[b]{0.45\textwidth}
    \includegraphics[width=\textwidth]{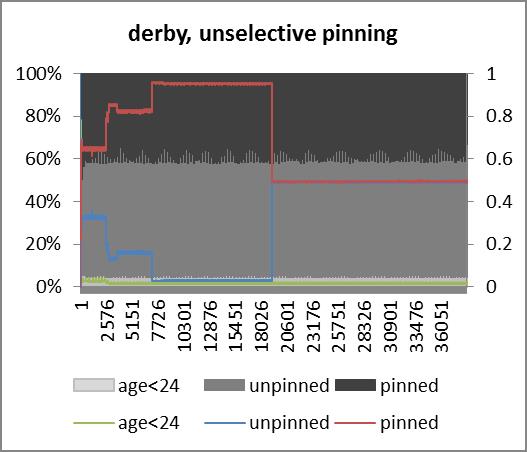}
    \caption{Derby, Unselective Pinning}
    \label{fig:3a}
  \end{subfigure}
  \begin{subfigure}[b]{0.45\textwidth}
    \includegraphics[width=\textwidth]{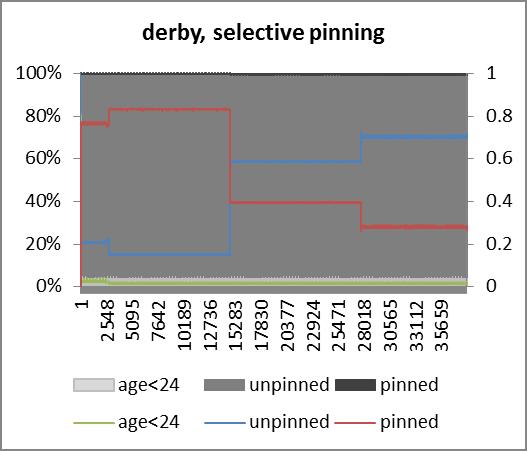}
    \caption{Derby, Selective Pinning}
    \label{fig:3b}
  \end{subfigure}
  
  \caption{Derby Activity}
  \label{fig:3}
\end{figure}

Most of the abrupt drops in reference activity in Figures~\ref{fig:3a} and \ref{fig:3b} coincide with cold collection, while increases tend to be associated with region pinning events. For the compiler.compiler benchmark most of the variability in region counts involved young regions. For the other benchmarks most of the variability involved unpinned regions. For all benchmarks the pinned region count was relatively stable, although more replacement occurred with unselective pinning. 

Ideally, pinned regions selection should result in a higher proportion of reference activity within pinned regions. Also, this should be realized by pinning as few regions as possible. By that measure, selective pinning outperformed unselective pinning for the compiler.compiler and xml.validation benchmarks and slightly underperformed for derby and xml.transform. 

The compiler.compiler workload mainly involves younger regions. It produced on average about 49 tenured regions, and most of these persisted for the duration of the run. With unselective pinning (Figure~\ref{fig:2a}) about 32 regions were typically pinned and they received about 33\% of reference activity on average. With selective pinning (Figure~\ref{fig:2b}) only 2 regions were typically pinned and they received about 45\% of reference activity on average.

The derby benchmark produced a greater and more stable population of  tenured regions, with a large number ($>$500) persisting over the course of the run. This is not surprising since derby loads an entire database into the heap before the benchmarking iteration starts and retains these objects for the duration of the run. 
With unselective pinning (Figure~\ref{fig:3a}) the maximum number P$_{max}$ of regions (256) were typically pinned at any time and they received about 69\% of reference activity on average. With selective pinning (Figure~\ref{fig:3b}) only 5 - 6 regions were typically pinned and they received about 53\% of reference activity on average.
If the P$_{max}$  limit had been removed for unselective pinning the number of pinned region would have risen to include more of the regions containing portions of the derby database content. This in turn would have reduced the number of unpinned regions available for compaction and tail filling, forcing allocation of new regions to receive aging heap objects. Selective pinning performed almost as well for derby with at most 6 pinned regions. 

\begin{figure}[ht!]
  \centering
  
  \begin{subfigure}[b]{0.45\textwidth}
    \includegraphics[width=\textwidth]{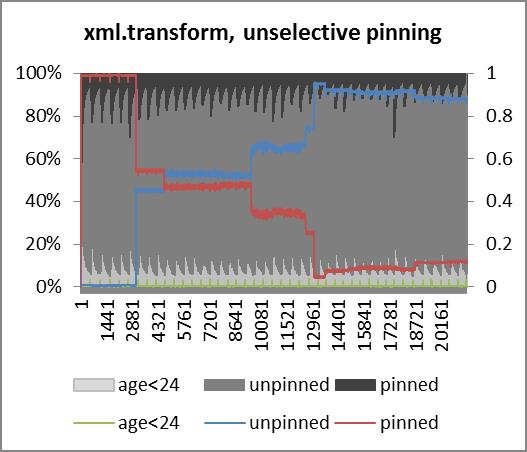}
    \caption{xml.transform, Unselective Pinning}
    \label{fig:4a}
  \end{subfigure}
  \begin{subfigure}[b]{0.45\textwidth}
    \includegraphics[width=\textwidth]{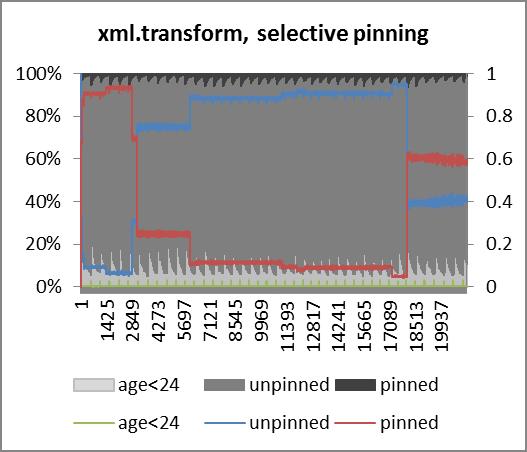}
    \caption{xml.transform, Selective Pinning}
    \label{fig:4b}
  \end{subfigure}
  
  \caption{xml.transform Activity}
  \label{fig:4}
\end{figure}

The xml.transform benchmark used an average of about 290 tenured regions, of which only about 40 persisted throughout the run. With unselective pinning (Figure~\ref{fig:4a}) about 23 regions were pinned, on average, and they received about 38\% of reference activity on average. With selective pinning (Figure~\ref{fig:4b}) only 6 regions were typically pinned and they received about 31\% of reference activity on average.

\begin{figure}[ht!]
  \centering
  
  \begin{subfigure}[b]{0.45\textwidth}
    \includegraphics[width=\textwidth]{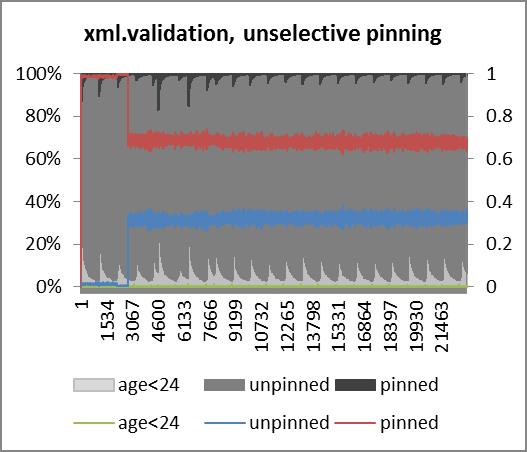}
    \caption{xml.validate, Unselective Pinning}
    \label{fig:5a}
  \end{subfigure}
  \begin{subfigure}[b]{0.45\textwidth}
    \includegraphics[width=\textwidth]{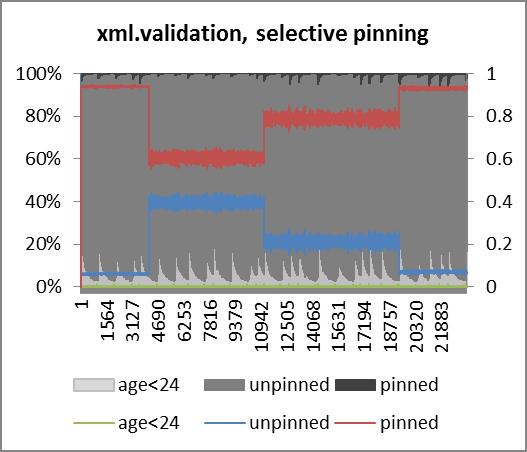}
    \caption{xml.validate, Selective Pinning}
    \label{fig:5b}
  \end{subfigure}
  
  \caption{xml.validate Activity}
  \label{fig:5}
\end{figure}

The xml.validation benchmark used about 200 tenured regions, with about 13 persisting for the duration of the run. With unselective pinning (Figure~\ref{fig:5a}) only about 7 regions were pinned at any time, and they received about 72\% of reference activity on average. With selective pinning (Figure~\ref{fig:5b}) only 2 regions were typically pinned and they received about 78\% of reference activity on average.

\subsection{Cold Object Collection}
Cold objects are collected into the cold area when their containing pinned regions pass a time (T$_{cold}$) where no new references into the region are sampled. The number and total size of cold objects collected into the cold area, and the number of references into the cold area, are summarized in Table \ref{tab_num_cold} for all runs of each SPECjvm2008~\cite{SPECjvm2008} benchmark profiled. The statistics for cold references include the total reference count and the number of distinct objects referenced. For all benchmark runs unselective pinning produced the greatest collection of cold objects, but it also tended to result in a higher count of references into the cold area, especially for xml.transform. In all cases, the number of distinct objects referenced was small, regardless of pinning strategy.

\begin{table}[!h]
\caption{The number and the size of cold objects}
\begin{center}
\begin{tabular}{| l | r | r | r | r |}
\hline
	&Cold Objects&Cold Bytes&\multicolumn{2}{|c|}{Cold References}\\	
\hline\hline
\multicolumn{5}{|l|}{compiler.compiler}\\
\hline
unselective, run1&24,452&4,498,952&	3&2\\
\hline
unselective, run2&15,717&8,009,752&0& \\
\hline
selective, run1&0&0&0& \\
\hline
selective, run2&0&0&0& \\
\hline
\multicolumn{5}{|l|}{derby}\\
\hline
unselective, run1&79,383&6,868,440&0& \\
\hline
unselective, run2&40,861&10,958,816&1&1\\
\hline
selective, run1&9,039&1,379,888&0& \\
\hline
selective, run2&3,284&279,736&0& \\
\hline
\multicolumn{5}{|l|}{xml.transform}\\
\hline
unselective, run1&27,603	&16,749,392&716,850&3\\
\hline
unselective, run2&29,995&17,394,688&635,768&2\\
\hline
selective, run1&14,486&8,928,248&0& 	\\
\hline
selective, run2&12,961&5,734,144&0& 	\\
\hline
\multicolumn{5}{|l|}{xml.validation}\\
\hline
unselective, run1&16,188&2,910,520&0&  \\
\hline
unselective, run2&14,926&2,733,280&	0& 	\\
\hline
selective, run1&3,698&474,904&0	&  \\
\hline
selective, run2&4,889&582,880&16&3\\
\hline\end{tabular}
\end{center}
\label{tab_num_cold}
\end{table}

\begin{figure}[ht!]
  \centering
  
  \begin{subfigure}[b]{0.45\textwidth}
    \includegraphics[width=\textwidth]{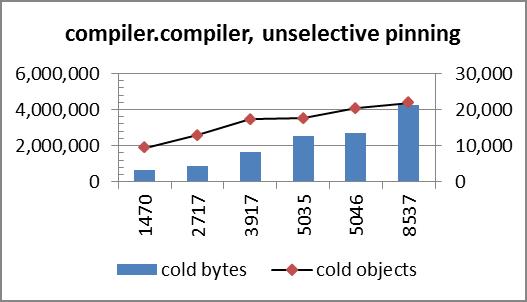}
    \caption{Compiler.compiler, Unselective Pinning}
    \label{fig:6a}
  \end{subfigure}
  \begin{subfigure}[b]{0.45\textwidth}
    \includegraphics[width=\textwidth]{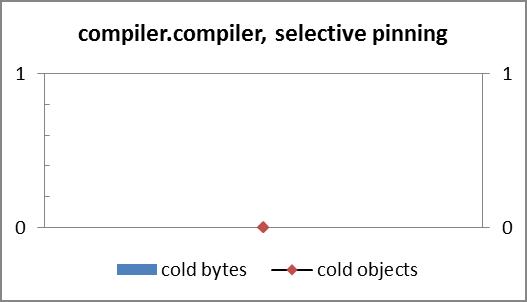}
    \caption{Compiler.compiler, Selective Pinning}
    \label{fig:6b}
  \end{subfigure}
  
  \caption{Compiler.compiler, Cold objects}
\end{figure}

Figures~\ref{fig:6a}-~\ref{fig:6b} show the cold collections for the first runs with unselective and selective pinning for each benchmark. The left axes represent total byte count; the right axes represent object count. Partial GC cycle counts at the time of cold collection are represented on the horizontal axes.

Unselective pinning for compiler.compiler resulted in a collection of over 30 pinned regions, six of which were cold collected. Three references to two cold objects were subsequently sampled. Selective pinning resulted in two regions that remained pinned for most of the compiler.compiler benchmark run. One of these went cold (no new activity for $>T_{cold}$ seconds) about halfway through the run and remained cold until the end but was not collectible because it's remembered card set was in a persistent overflow state.

\begin{figure}[ht!]
  \centering
  
  \begin{subfigure}[b]{0.45\textwidth}
    \includegraphics[width=\textwidth]{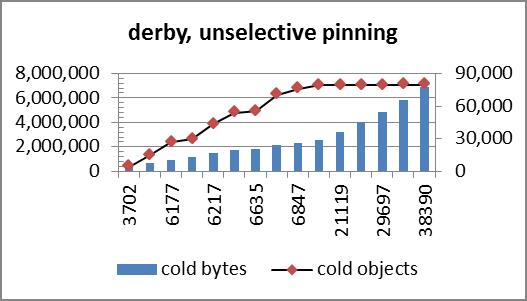}
    \caption{Derby, Unselective Pinning}
    \label{fig:7a}
  \end{subfigure}
  \begin{subfigure}[b]{0.45\textwidth}
    \includegraphics[width=\textwidth]{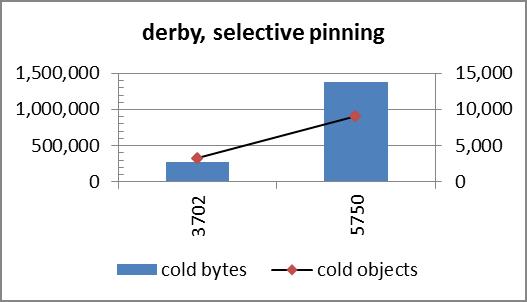}
    \caption{Derby, Selective Pinning}
    \label{fig:7b}
  \end{subfigure}
  
  \caption{Derby, Cold objects}
  \label{fig:7}
\end{figure}

Unselective pinning for derby resulted in a maximal collection of pinned regions (256 regions) and 15 cold collections. No activity was recorded in the cold area. Selective pinning pinned only six very active regions, two of which were cold collected early in the run. There was no activity in the cold area. 

\begin{figure}[ht!]
  \centering
  
  \begin{subfigure}[b]{0.45\textwidth}
    \includegraphics[width=\textwidth]{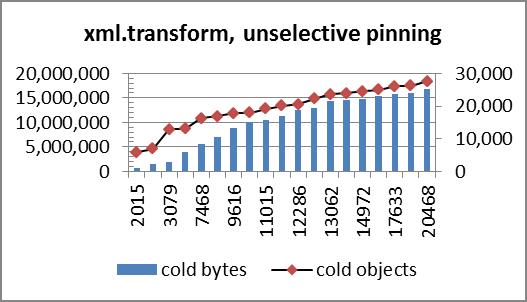}
    \caption{xml.transform, Unselective Pinning}
    \label{fig:8a}
  \end{subfigure}
  \begin{subfigure}[b]{0.45\textwidth}
    \includegraphics[width=\textwidth]{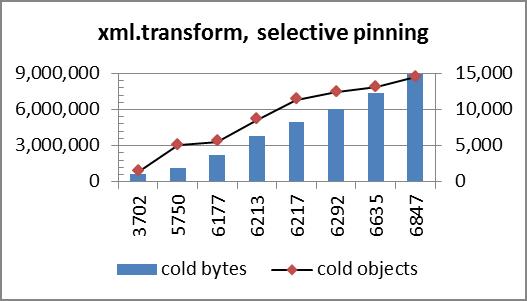}
    \caption{xml.transform, Selective Pinning}
    \label{fig:8b}
  \end{subfigure}
  
  \caption{xml.transform, Cold objects}
  \label{fig:8}
\end{figure}

Unselective pinning for xml.transform resulted in a collection of about 23 pinned regions, 19 of which were cold collected. However, there were a high number of references to three objects in the cold area. Selective pinning pinned at most six active regions at any time but eight regions were cold collected. There was no subsequent activity in the cold area. 

\begin{figure}[ht!]
  \centering
  
  \begin{subfigure}[b]{0.45\textwidth}
    \includegraphics[width=\textwidth]{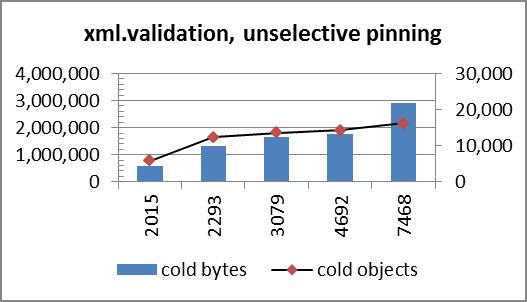}
    \caption{xml.validate, Unselective Pinning}
    \label{fig:9a}
  \end{subfigure}
  \begin{subfigure}[b]{0.45\textwidth}
    \includegraphics[width=\textwidth]{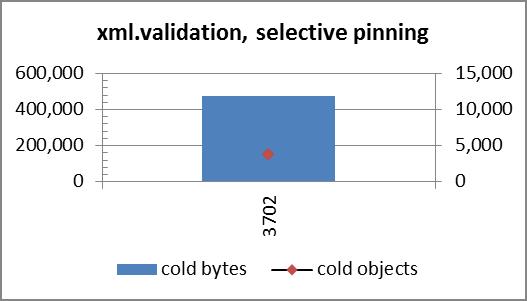}
    \caption{xml.validate, Selective Pinning}
    \label{fig:9b}
  \end{subfigure}
  
  \caption{xml.validate, Cold objects}
\end{figure}

Unselective pinning for xml.validation resulted in a collection of about seven pinned regions, five of which were cold collected. There were no references to objects in the cold area. Selective pinning pinned at most two active regions at any time and cold collected one region. There was no subsequent activity in the cold area.

\section{Evaluation of the Stack-Based Solution with an Access Barrier}
Since stack sampling is intermittent, with each mutator thread walking its stack about once per millisecond, and can only occur at safe points only, there is a concern that the reference sampling rate may not be high enough to support reliable cold object identification. An Access Barrier can capture all read/write access operations when Java runs interpreting mode. Since the Access Barrier does not miss any access information, it will be used as benchmark to evaluate the correctness and efficiency of stack-based solution.

\subsection{Evaluation Metrics}
Two key metrics are used to evaluate stack-based solution. $FalseInactivity$ is used to verify the reliability of stack-based solution.  $ConvergenceTime$ is used to evaluate the efficiency of stack-based solution.

\begin{itemize}
\item $FalseInactivity$
is the number of objects that are considered inactive with the stack-based solution, but are marked active with the Access Barrier.
For example, because of non-continuous sampling, the stack-based solution misses some objects' activities, and these objects are considered to be inactive. 
However, these objects' activities are captured by the Access Barrier.
The $FalseInactivity$ reflects the missing of some active objects. 
Smaller values mean better stack-based solutions.
The $FalseInactivity$ is a ratio described in the following formula.

\begin{equation}
   FalseInactivity = \frac{Numbers\_of\_false\_inactive\_objects} {Numbers\_of\_all\_inactive\_objects}   
\end{equation}

\item $ConvergenceTime$
is the  time span that a region is pinned before it is determined to have identified all active objects.  $ConvergenceTime$ reflects the speed that a pinned region is found to collect cold objects. The lower the $ConvergenceTime$ is, the more efficient the identification of cold objects is.

\end{itemize}

\subsection{Experiments}
Two evaluation experiments are performed with Java executing in interpreting mode instead of Just-in-Time mode. 

\subsection{Case 1 - SPECjvm2008 Derby}
Running parameters are as follows.

\begin{enumerate}
\item Running period: 60 hours
\item Cold threshold: 6 hours
\item Sampling interval: 100 $ms$
\end{enumerate}

Benchmark Derby was run for 60 hours, active objects were identified with stack-based and Access Barrier at the same time. Experimental results are presented in Table~\ref{tab_eval}. It is not surprising that the Access Barrier can harvest more cold objects than the stack-based solution, because the Access Barrier can capture all read/write access to objects, while stack sampling is intermittent. For example, the number of collectible pinned regions in Access Barrier is 5 times larger than in stack-based solution; the number of cold objects in Access Barrier is 11.78 times larger than that in stack-based solution; and the size of cold objects is 10.42 times larger in Access Barrier than that in stack-based solution.

\begin{table}[!h]
\caption{Evaluation with SPECjvm2008 Derby}
\begin{center}
\begin{tabular}{|l|r|r|c|}
\hline
Items&AccessBarrier&stack-based&Ratio\\
\hline\hline

Collectible pinned regions&85&17& 5 : 1\\
\hline
CovergenceTime(in Second)&27,721.62& 69,497.71&1:2.50 \\
\hline
All Objects&1,485,531&72,290& \\
\hline
Active Objects&673,272& 3,350& \\
\hline
Cold Objects&812,259&68,940& 11.78:1\\
\hline
Size of All Objects(in Byte)& 172,021,784 &32,882,056  & \\
\hline
Size of Active Objects(in Byte)&129,221,416&28,773,560 & \\
\hline
Size of Cold Objects(in Byte)& 42,800,368&4,108,496 &10.42:1\\
\hline
\end{tabular}
\end{center}
\label{tab_eval}
\end{table}

\subsubsection{Reliability of the stack-based solution.}
The $FalseInactivity$ ratio is 1.62\% (see Table~\ref{tab_num_cold}), which is quite low. It reflects the fact that there are few objects that are incorrectly classified as cold. The data supports the hypothesis that cold objects can be identified by the stack-based approach, which is encouraging. 

\begin{table}[!h]
\caption{$FalseInactivity$ Results }
\begin{center}
\begin{tabular}{| c || c | c |}
\hline
Inactive objects & $FalseInactivity$ Objects & $FalseInactivity$ Ratio\\
\hline\hline
 68,940 &1,117&1.62\%\\
\hline
\end{tabular}
\end{center}
\end{table}

\subsubsection{Efficiency of the stack-based marking approach.}
The Access Barrier has found 85 collectible pinned regions, while the stack-based solution has 17 collectible pinned regions. Although the Access Barrier has more collectible pinned regions than the stack-based solution, 17 collectible pinned regions in the stack-based solution are completely included in the Access Barrier collectible pinned regions. 

\begin{figure}[ht!]
\centering
\includegraphics[width=120mm, height=55mm]{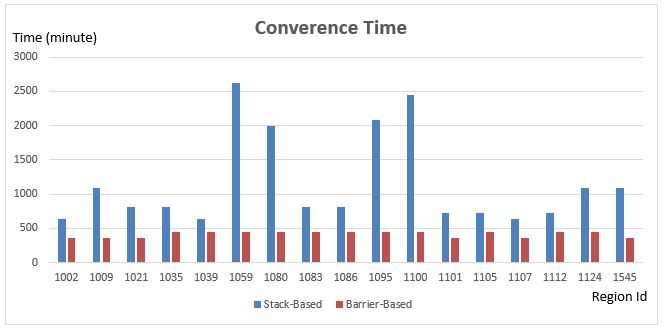}
\caption{Convergence time in Derby}
\label{fig:10}
\end{figure}

Figure~\ref{fig:10} shows a convergence time comparison between stack-based solution and the Access Barrier. The X-axis represents 17 common collectible pinned regions, the Y-axis represents convergence time. In the Access Barrier, the maximum convergence time is less than 500 minutes, while in stack-based solution, the maximum convergence time reaches more than 2500 minutes.

\subsection{Case 2 - SPECjvm2008 Compiler.compiler}
Running parameters are as follows.

\begin{enumerate}
\item Running period: 12 hours
\item Cold threshold: 72 minutes
\item Sampling interval: 15 $ms$
\end{enumerate}

After Compiler.compiler has run by 12 hours, the results shown in Table \ref{tab_eval_compcomp} are obtained. The Access Barrier still harvests more cold objects than the stack-based solution.

\begin{table}[!h]
\caption{Evaluation with Compiler.compiler}
\begin{center}
\begin{tabular}{| l || r | r | l |}
\hline
Items&AccessBarrier&stack-based&Ratio\\
\hline\hline

Collectible pinned regions&64&28& 2.29 : 1\\
\hline
AverageColdDuration (in Seconds)&4440.64&6311.96&1 : 1.42 \\
\hline
All Objects&1,390,739& 376,824 & \\
\hline
Active Objects&141,772&279& \\
\hline
Cold Objects&1,248,967&376,545& 3.32 : 1\\
\hline
Size of All Object (in Bytes)&3.32&53,923,840& \\
\hline
Size of Active Object (in Bytes)&20,090,576&1,280,616& \\
\hline
Size of Cold Object (in Bytes)&109,475,216&52,643,224&2.08 : 1\\
\hline
\end{tabular}
\end{center}
\label{tab_eval_compcomp} 
\end{table}

\subsubsection{Reliability of stack-based solution.}
The falseInactivity ratio is 0.32\% (see Table~\ref{tab_false}), which is still quite low. The data confirms  that cold objects can be identified by the stack-based approach. 

\begin{table}[!h]
\caption{$FalseInactivity$}
\begin{center}
\begin{tabular}{| c || c | c |}
\hline
Inactive objects& $FalseInactivity$ Objects&falseInactivity Ratio\\
\hline\hline
376,545&1,191&0.32\%\\
\hline
\end{tabular}
\end{center}
 \label{tab_false}
\end{table}

\subsubsection{Convergence time analysis.}
The Access Barrier has 64 collectible pinned regions. The stack-based solution has 28 collectible pinned regions. Although the Access Barrier has more collectible pinned regions than the stack-based solution, 28 collectible pinned regions in the stack-based solution are completely included in the Barrier-based collectible pinned regions. 

\begin{figure}[ht!]
\centering
\includegraphics[width=120mm, height=55mm]{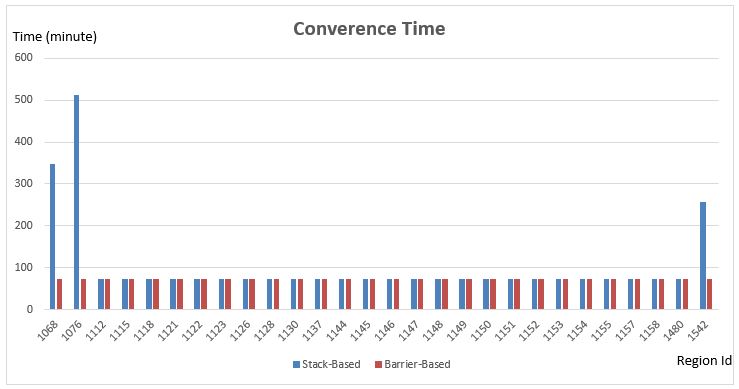}
\caption{Convergence time in Compiler.compiler}
\label{fig:11}
\end{figure}

Figure~\ref{fig:11} shows a convergence time comparison between stack-based solution and Access Barrier. In the Access Barrier, the maximum convergence time is less than 100 minutes. While in stack-based solution, the convergence time in the majority of collectible pinned regions is less than 100 minutes as well, only 3 collectible pinned regions have a high convergence time.

\section{DISCUSSION}
During the ssd-stack JVM benchmarking runs, mutator threads walked their stacks to harvest heap references once every 1-2 $ms$ on average. For cold object identification to be reliable and effective, the rate of reference sampling must be such that any active object within a pinned region is likely to appear on a mutator stack at a stack walking safe point at least once while the region is pinned. The $FalseInactivity$ experimental results in Derby and Compiler.compiler show that the sampling frequency with 1-2 $ms$ can satisfy the requirement of cold object identification.

A very small number of cold objects were referenced during any of the ssd-stack JVM runs, and with the exception of xml.transform cold objects were referenced very infrequently. Most benchmark runs collected a few tens of thousands of cold objects. In the exceptional case of compiller.compiler with selective pinning, one pinned region went cold after 1,757 partial GC cycles but was not collectible due to a persistently overflowed remembered set; none of the 3,354 objects that were cold at that point received references for the remainder of the benchmark run (for 8,055 subsequent partial GC cycles).

The ssd-stack JVM, with selective or unselective pinning, consistently resulted in increased memory pressure, timeslicing, and kernel CPU usage compared to the linux JVM. The singleton thread activity sampling daemon minimizes writes to pinned region activity maps but must test the activity bit for every sampled reference into a pinned region, making activity maps high running candidates for available cache lines. Although cache misses were not profiled for these runs, it is likely that high frequency access to pinned region activity maps from the thread activity sampling daemon had a significant effect on memory bandwidth. Since the daemon thread is bound to a specific node this effect may be limited, especially in larger multicore systems. However, the daemon does present a multicore scalability problem since it must handle proportionately larger loads as the number of available cores increases. 

Most of the ssd-stack benchmarking runs yielded a few tens of megabytes in the cold area and consumed 4-6\% of the available CPU bandwidth, which is a relatively high price to pay for the amount of cold data collected. In this paper, only primitive arrays and leaf objects are considered as cold objects. If cold objects are not limited to primitive arrays and leaf objects, the amount of cold data collected should be increased, but the possibility of object references from cold objects to active objects would then require that marking should traverse into the cold area, which is undesirable.

The region pinning framework attempts to pin the most active regions because they highlight mutator activity and allow cold objects to be identified more quickly and with greater confidence. Selective pinning sets a high bar on the activity metric for selectable regions and tends to pin only a few regions, without replacement. It tended to collect a relatively small but very stable set of cold objects. Unselective pinning attempts to maximize the number of pinned regions and selects the most active regions in batches, but tends to unpin and replace these over time. This strategy produced more substantial cold collections and the cold area typically received a small amount of reference activity that was confined to a small number of distinct objects.

\section{CONCLUSION}
In this paper, we show the stack-based cold object identification framework, which samples the mutator thread stack, marks the active objects, and harvests the cold objects. stack-based reference sampling was effective in identifying inactive objects for the SPECjvm2008~\cite{SPECjvm2008} benchmarks studied here, as evidenced by the stability of the cold areas established during the benchmark runs. A few tens of megabytes of cold objects have been identified and harvested into cold regions. Furthermore, we evaluate the correctness and efficiency of stack-based solution with an Access barrier implementation, the results support that the stack-based solution is an acceptable cold object identification approach.

The runtime overhead for walking mutator stacks and maintaining pinned activity maps offset any gains that accrued from establishing the cold area and marshalling cold objects out of resident memory, but there is still space to reduce overhead by optimization. 

\section{FUTURE WORK}
The focus of this effort so far has been on cold object identification and sequestration. If further development is extended, a mechanism for managing frequently active objects in the cold area should be developed. For example, if madvise() \cite{Mckusick86anew} is used to sequester pages in the cold area, pages containing parts of active objects can be excluded. This would be effective if these objects are relatively rare and simpler than providing specialized methods to copy active cold objects back into tenured regions. 

Only limited amount of work has been done to verify that the JVM and GC do not reference objects in the cold area. Some JIT and GC (concurrent mark) activity in cold areas was detected by erecting a partial memory protection (read/write) barrier around the cold area between GC cycles. A similar mechanism can be used to detect GC incursions into the cold area during GC cycles, but this has not been investigated to date. The sources of these incursions will need to be modified to suppress activity in the cold area. For example, objects being marked in the root set, where the leaf marking optimization is not available, can be tested for inclusion in cold regions and treated as leaf objects in that case. The JIT peeks into cold regions to determine the length of arrays but it may be possible to apply defaults or forego optimizations for arrays located within the cold area.

\section*{Acknowledgement}
 The authors would like to acknowledge the funding support provided by IBM and the Atlantic Canada Opportunities Agency (ACOA) through the Atlantic Innovation Fund (AIF) program. Furthermore, we would also like to thank the New Brunswick Innovation Fund for contributing to this project. Finally, we would like to thank the Centre for Advanced 
Studies - Atlantic for access to the resources for conducting our research.

\bibliographystyle{plain}
\bibliography{GC}

%
%
%

\end{document}